# Empirical evidence for a global atmospheric temperature control system: physical structure

L. Mark W. Leggett* & David A. Ball

Global Risk Policy Group Pty Ltd, Townsville, Queensland, Australia

*mleggett.globalriskprogress@gmail.com

29 September 2020

**Abstract**

There is evidence that a natural control system influences global atmospheric surface temperature (Leggett and Ball, 2020). The present paper sets up and tests a hypothesis concerning the physical makeup of the sequential elements of the control system and its outcomes. The final outcome the control system influences is defined as global atmospheric surface temperature. The terms used for the control system element types in the hypothesis are, in sequence: leading element, controller and actuator. Actuators are hypothesised to affect, in turn, the final outcome: either directly, or via penultimate outcomes. The existence of the control system is evidenced by demonstration of statistically significant one-way Granger causality across each step of the hypothesised control system sequence. Evidence is presented that the leading element of the control system, represented by the Normalized Difference Vegetation Index, is the global biosphere. The biosphere as a control system has previously been referred to as Gaia (Lovelock and Margulis, 1974). A fourth, second-derivative, term is found to enhance the Proportional, Integral and Derivative process terms of the control system shown in Leggett and Ball (2020). The main actuators of the control system found are shown to be wind speed and cloud cover. Cloud cover is shown to influence the final outcome, global surface temperature, directly. It and wind speed also influence the penultimate outcomes found, those of enhanced ocean heat uptake and enhanced outgoing longwave radiation. These together lead to control system output to the final outcome, global atmospheric temperature. Overall, evidence for the activity of the control system is present across many major physical dimensions of the Earth's atmosphere.

**1. Introduction**

In Leggett and Ball (2020), we provided statistically significant observational evidence that a feedback control system moderating atmospheric temperature is presently operating coherently at global scale. Further, this control system was shown to be of a sophisticated type, displaying controller process terms with proportional-integral-derivative characteristics.

This paper provides further evidence about the controller process terms of the control system, and also about its physical components.

## 1.1 Terminology

After Sahib (2015), we use the expression 'n-term control type' to describe controller process terms. Hence, the Proportional-Integral-Derivative (PID) control type is classed as a 'three-term control type'.

In describing the components of a physical control system, Åström and Murray (2008) write:

> A modern controller senses the operation of a system, compares it against the desired behavior, computes corrective actions based on a model of the system's response to external inputs and actuates the system to effect the desired change. This basic feedback loop of sensing, computation and actuation is the central concept in control.

Accordingly, we will group candidate controller process terms and candidate physical components for the control system under the headings 'controller' and 'actuator' respectively. These make up 'elements' of the control system.

Noting that physical systems with time-varying internal couplings between components are abundant in nature, BozorgMagham et al. (2015) observe that the definitive approach to detecting causal relationships between such elements of a system is to fully identify the underlying mechanisms. They note, however, that observed data is required to establish causal relationships between candidate elements in the absence of complete knowledge of internal components

In this paper, we use observed data to seek causal relationships between candidate elements of the control system. Time series analysis is used for this purpose.

Two major methods for seeking causal relationships are Granger causality (Granger, 1969) and Convergent Cross Mapping (CCM) (Sugihara et al., 2012; BozorgMagham et al., 2015). Each provides a framework that uses predictability, as opposed to correlation, to identify causation between time-series variables (Sugihara et al., 2012).

Use of Granger causality or CCM in connectivity studies also enables identification of the *directionality* of causal interaction or information flow between the putative components of a control system (for Granger causality, see Granger, 1969; for CCM, Sugihara et al., 2012). CCM is relevant when there is major phase shifting over time between variables, such that Granger causality testing would not show a relationship (Sugihara et al., 2012).

Granger causality was used in initial data analysis for this study to determine whether or not CCM would be required. This was found not to be the case, and therefore Granger causality was used for this study.

A time series $X$ is said to be Granger-causal of $Y$ if it can be shown that values of $X$ provide statistically significant information about future values of $Y$ that is not contained in past values of $Y$ and other relevant information (Kaufmann and Stern, 1997; Terrell, 2019). If Granger causality is directional, information from *Y* is not present in future values of *X*.

We note that the evidence for directional Granger causality between *X* and *Y* is in the domain of information, implying information flow between *X* and *Y*. In this paper we sometimes use the term 'information flow' instead of 'causality'. If *X* is Granger causal of *Y* we sometimes describe this as Granger causal connectivity.

We use the term 'one-way Granger causality' to describe the directionality of causality or information flow.

In this study, Granger causality in its form as an output of Vector Autoregression (VAR) analysis (Sims, 1980; Stock and Watson, 2001) was used to assess relationships between candidate control system elements. A relationship is said to be Granger causal if the probability of it being due to chance is less than a specified value. The probability level used is this study is the standard 0.05 statistical significance level.

As well as its probability, the *strength* of Granger causality is also of interest, and a method of estimation has been developed (Geweke, 1982; Hesse et al., 2003). The method estimates the strength of Granger causality, for example of *Y* to *X*, as a measure of linear feedback between two signals. This estimate is made for each time period or lag assessed. The maximum value found from this series of *Y* to *X* estimates represents a simple measure for the strength of feedback in the *Y* to *X* direction.

In this paper, the strength of Granger causal interaction in both the *Y* to *X* and the *X* to *Y* directions is calculated according to Geweke's equations 4.1 to 4.4 (Geweke, 1982, p.209); and Hesse's equations 5 to 7 (Hesse et al., 2003). The strength of causality is termed $F_{YX}$ and $F_{XY}$ respectively here. The calculations were conducted using code (available on request) written for the statistical analysis package used for this study (IHS EViews, 2017).

Where climate studies are concerned, Hlinka et al. (2017) note that the majority of climate network studies have been limited to the use of symmetric dependence measures such as correlation and recommends an approach to climate network studies based on prediction not correlation and which enables the capture of the directionality of causal interaction.

Some relationships were assessed using multiple regression of the autoregressive distributed lag (ARDL) form (Pesaran et al., 2001). Section 2 (Methods) provides further information on the approach to VAR and ARDL analysis used in this study.

**1.2 Hypothesis**

With the above background, we hypothesise a control system with putative elements A, B, C, D, etc. that make up a sequential element chain. If Granger causality tests show that A causes B but not the reverse, and the same is true for B to C, C to D, and so on, we have determined that the elements unambiguously make up the causal chain as hypothesised. This we term 'end-to-end one-way Granger causal connectivity'.

For the element at the start of the control system sequence, we use the term 'leading element' (BozorgMagham et al., 2015). The next set of elements we term 'controllers', the following set we term 'actuators', which may then directly affect the 'outcome' or control what we term 'penultimate outcomes' and then affect the final outcome. The penultimate outcomes of the control system are defined here as intermediate outcomes of the system which then contribute to the final outcome.

There are many drivers of atmospheric trends (IPCC, 2013; Stern and Kaufman, 2014). We note that the elements for which we are seeking evidence in the control system do not need to be the largest driver of the subsequent link in the control system chain in order to contribute to an effective control system.

With the above background, we hypothesise the existence of a global atmospheric control system displaying end-to-end one-way Granger causal connectivity in a chain from control system element to subsequent control system element, starting from the leading element. The hypothesis is depicted diagrammatically in Table 1.

Table 1. Hypothesised flow of causality by control system element across the hypothesised causal chain

| Control system step | From | Flow of Causality | To |
| --- | --- | --- | --- |
| 1 | Candidate leading element | ➔ and not ⬅ | Candidate controller terms |
| 2 | Candidate controller terms | ➔ and not ⬅ | Candidate actuators |
| 3 | Candidate actuators | ➔ and not ⬅ | Outcome |

### 1.3 Possibility of a multi-channel control system

We note that the multiple, three-term, PID control type shown in Leggett and Ball (2020) enables the possibility that a multi-channel control system exists and this is explored for.

### 1.4 Link-to-link and longer-range analyses

The study aim is to search for evidence of causal chains of physical control system stages by seeking one-way causality from each stage of such putative causal chains to the next, using Granger causality analysis.
.

The hypothesised chain in Table 1 has three stages. Let us term the stages A, B, and C.

An issue may potentially arise in the link-to-link analysis of this chain, in that it may provide strong evidence for connectivity from A to B but the evidence for connectivity between B and C may exist, but may be less strong.

In this situation one can turn from link-to-link studies of connectivity to the presence of information at longer range. Demonstrating strong connectivity from A to C would indicate that the B to C issue represents either a possible data issue with the B to C connection, or that an additional connection exists other than the one being tested in this study.

This method of investigation has much in common with the use of electronic signal tracer test equipment used to troubleshoot radio and other electronic circuitry (Bishop, 2007). Here, a test signal is injected into the device under test. Then, by using the signal tracer, the signal can be followed through the various circuits of the device. So long as the signal can be detected, the circuitry up to that point is (at least minimally) connected. If the signal is not detected there is no evidence of a connection between those circuit elements.

Such longer-range causality analyses are used from time to time in this study.

### 1.5 Protocol for each Granger causality analysis

The study investigates the existence of multiple control system channels, each with multiple chains of elements and associated causal steps. This is done by carrying out a separate Granger causality analysis for each pair of elements or steps,

As a result, many Granger causality studies are reported in this paper. The *probability* of Granger causality is the measure reported in the majority of Granger causality studies, and for reasons of space is the main measure used here.

That said, the *strength* of Granger causality is also assessed in some situations. The reason for this is as follows.

In assessing the existence of putative control system steps in line with the hypothesis for one-way Granger causality between any two candidate elements, the following types of results may occur:

> 1. No statistically significant Granger causality is found between the two elements. In this case the result is reported and no further assessment carried out.
>
> 2. Statistically significant Granger causality is found for a control system step, but it applies in both directions. This result it is not treated as evidence for the control system.

> 3. One-way Granger causality in the proposed control system direction is found based on the 0.05 significance level, and the evidence for causality in the other direction is well in excess of the 0.05 level (greater than 0.1). Results of this type are considered sufficient to meet the control system hypothesis.
>
> 4. Evidence is as for category 3 except that the evidence for reverse causality is greater than, but close to, the 0.05 level (between 0.05 and 0.1). Here two-way causality is not being strongly rejected. In this situation a second line of evidence is sought, using strength of Granger causality.

In this last instance, if the result of the analysis of the strength of Granger causality for the putative control system step is stronger in the direction of the control system hypothesis than in the other direction, then these two pieces of evidence are considered to provide strong evidence overall that the putative control system step is likely to exist.

### 1.6 Note on the scope of the study

The scope of this paper is limited to seeking evidence that supports the hypothesis. As such, potential connections between many pairs of climate series are assessed. Some connections may not support the hypothesis, but are still reported. While further investigation of these may be of some interest, they are generally not discussed further in this paper in order to maintain the focus on the hypothesis.

## 2. Methods

Statistical methods used are standard (Greene, 2012) and generally as used in Leggett and Ball (2015; 2018). Categories of methods used are: normalisation; differentiation (approximated by differencing); integration (approximated by the cumulative sum); and time-series analysis.

Within time-series analysis, methods used are: *Z*-scoring; smoothing; testing for the order of integration of each series (a prerequisite for using data series in time-series analysis); autoregressive distributed lag (ARDL) dynamic regression modelling to include autocorrelation in models; and Vector Autoregression (VAR) modelling to enable Granger causality testing. These methods will now be described in turn.

### 2.1 Normalisation of data series

To make it easier to visually assess the relationship between the variables and to estimate the relative scale of regression coefficients in multiple regression analysis, each data series was normalised using statistical *Z* scores (also known as standardised deviation scores). In a *Z*-scored data series, each data point is part of an overall data series that sums to a zero mean and variance of 1, enabling comparison of data having different native units. Hence, when several *Z*-scored time series are depicted in a graph, all the time series will closely superimpose, enabling visual inspection to clearly discern the degree of similarity or dissimilarity between them. Individual figure legends contain details on the series lengths used as base periods for the *Z*-scoring.

A regression using *Z*-scored variables provides standardised regression coefficients. These coefficients report how much change a one-standard-deviation change in the independent variable produces in the dependent variable. Although comparisons between these coefficients must be interpreted with care, a standardised coefficient for independent variable *a* of 2, for example, indicates that independent variable *a* is twice as influential upon the dependent variable as another independent variable that has a standardised coefficient of 1 (Allen, 1997).

**2.2 Smoothing of data**

Smoothing was used on some data series to the extent required to reveal if significant relationships were present. Smoothing using moving averages removes large month-to-month variations to reveal the underlying pattern or trend. Smoothing was carried out by means of a 13-month moving average, which also minimises seasonal effects.

**2.3 Labelling data series**

A key explaining the labels used for each series in the paper, including its smoothing and *Z*-scoring status, is provided in Table 3 in the Supplementary Information.

**2.4 Notation for candidate control elements**

Given that the present disturbance to global temperature is leading to an increase, and this disturbance is resisted by the control system, any candidate control element would be expected to show a decreasing trend (Leggett and Ball, 2020).

As these series of themselves are often increasing, they are converted to control system terms by reversing their sign. Each candidate control series is identified as such by the suffix '_CONTR'.

Two further series are reversed, but are considered to be penultimate outcomes, not candidate control elements. These are given the suffix ' _REVERSE'.

A regression model supporting the hypothesis that the temperature series correlates with one or more control elements will display a minus sign for each supported element.

For readability in text, the suffixes may be omitted and the word order of the descriptor changed. For example Cloud_Ocean_CONTR may be written 'Ocean cloud' or 'Ocean cloud cover'. Sign reversal does not affect the operation of Granger causality tests.

**2.5 Time series analysis**

Time series models differ from ordinary regression models in that the results are in a sequence. Hence, the dependent variable is influenced not only by the independent variables, but also by prior values of the dependent variable itself. This is termed autocorrelation between measured values. The serial nature of the measurements must be addressed by careful examination of the lag structure of the model. This type of ordinary least squares regression is termed 'time series analysis' (Greene, 2012).

A further issue in time series analysis concerns what is termed the 'order of integration' of each of the series used. Greene (2012) states: "The series yt is said to be integrated of order one, denoted I(1), because taking a first difference produces a stationary process. A non-stationary series is integrated of order d, denoted I(d), if it becomes stationary after being first-differenced d times. An I(1) series in its raw (undifferenced) form will typically be constantly growing, or wandering about with no tendency to revert to a fixed mean."

In this paper, from time to time we deal with combinations of series with different orders of integration. Methods used to address this are included in the following sections.

**2.5.1 Regression**

For regression, the ARDL method (Pesaran et al., 2001) can be used whether variables are purely of order of integration I(0) (a stationary series), purely I(1) or a mixture of both I(0) and I(1) (Greene, 2012; Janjua et al., 2014; Ahmad and Du, 2017).

For ARDL, the stationarity or otherwise of each series must still be assessed to ensure that there are no I(2) or higher series present. ARDL does not work for such series (Pesaran et al., 2001). A range of tests exists to assess stationarity. In this study, the augmented Dickey-Fuller (ADF) test is used.

The usual tests for stationarity (for example, the ADF test) can allow for the presence of a *linear* deterministic trend in the time-series in question, but they cannot allow for the possibility of a *polynomial* trend. It will be shown that this possibility needs to be tested for in this study.

There is evidence that if a unit root test that allows for only a linear trend is applied mistakenly to a series with a polynomial trend, then the test may have extremely low power. For example, see the results of Harvey et al. (2008) in the context of the de-trended ADF-type tests of Elliott et al. (1996).

This problem may be resolved by using the Lagrange multiplier (LM) unit root test(s) proposed by Schmidt and Phillips (1992). These tests make explicit allowance for the possibility of a polynomial deterministic trend of order up to 4 in the series under test. As with the ADF test, the null hypothesis is that the series has a unit root, and the alternative hypothesis is that it is stationary. A value for the LM test statistic that is more negative than the tabulated critical value leads to a *rejection* of the null hypothesis, and suggests that the series is stationary.

The Schmidt-Phillips test is available in the 'urca' package in the R statistical software (R Development Core Team, 2009) and is described in the documentation by Pfaff et al. (2016). Specifically, the ur.sp-class allows us to apply the Schmidt-Phillips tests. Although there are two such tests, the so-called $\tau$ test and the $\rho$ test, the results of only the former test are reported (The results of the $\rho$ test in our models led to exactly the same conclusions). Testing is reported at the 1% significance level, but the results are not sensitive to this choice.

The order of integration of each series used has been determined and is presented in Tables 1 and 2 in the Supplementary Information. The tables show that each series displays an order of integration suitable for use in the ARDL multiple regressions and VAR models reported.

The ARDL method uses special significance tests (called 'bounds tests') (Pesaran et al., 2001). These test the significance of results against both an I(0) realm and an I(1) realm. If the result passes the test for each realm, a defensible model is obtained.

If the outcome of the bounds testing is positive, a long-run 'levels model' is estimated. These results are then used to measure short-run dynamic effects.

The above is carried out for models in which any autocorrelation present in the short-run relationship is fully accounted for by use of an optimal lag structure. In this study, this is done within the modelling process by reference to the adjusted coefficient of determination, $R^2$, and to the Akaike Information Criterion.

The adjusted $R^2$ is used because there is a problem that the unadjusted $R^2$ cannot fall when independent variables are added to a model in multiple regression (Greene, 2012). To address this issue, the adjusted $R^2$ is a fit measure that penalises the loss of degrees of freedom that result from adding variables to the model. There is, however, some question about whether the penalty is sufficiently large to ensure that the criterion will necessarily lead the analyst to the correct model as the sample size increases (Greene, 2012). Several alternative fit measures termed 'information criteria' have been developed. Several criteria are available for selection within the Eviews ARDL method. We use the ARDL default criterion, the Akaike Information Criterion.

Pesaran et al. (2001) point out that ARDL modelling is "also based on the assumption that the disturbances … are serially uncorrelated. It is therefore important that the lag order p of the underlying VAR is selected appropriately. There is a delicate balance between choosing p sufficiently large to mitigate the residual serial correlation problem and, at the same time, sufficiently small so that the conditional ECM is not unduly over-parameterized, particularly (when) limited time series data … are available."

To avoid over-parameterisation, we seek the model with the fewest lags that produces a model with no autocorrelation affecting results.

The specific information sought from each well-specified ARDL model is the effect on the dependent variable of each of the potential control variables, and the degree of statistical significance of each effect.

We note that, to avoid duplication of results, in the ARDL analyses we focus on the long-run model results; shorter dynamics, which are at the heart of VAR Granger causality analysis, are dealt with there.

The ARDL econometric model used here was that implemented in the Eviews 10 package (IHS Eviews, 2017). The time series statistical software package Gnu Regression, Econometrics and Time-series Library (GRETL) was also used (accessed 17 June, 2019).

### 2.5.2 Granger causality

We have previously noted (Leggett and Ball, 2015), that Granger causality (Granger, 1969) more closely approaches the quality of information derived from random placement into experimental and control categories, although using correlational data.

According to Stern and Kaufmann (2014), a time-series variable B$x$ (e.g. atmospheric $CO_2$) is said to be Granger-causal of variable B$y$ (e.g. surface temperature) if past values of $x$ help to predict the current level of $y$ better than just the past values of $y$ do, given all other relevant information. Granger causality has been noted as being quite close to information theory (Amblard and Michel, 2013). The case where information is being transmitted against a background containing a noise – or stochastic – element is central to information theory. Bearing the noise element in mind, if some of the signature ('signal') of series $a$ is seen at a later point in the makeup of the signal of series $b$ (and the opposite is not true), this becomes the focus.

As in Leggett and Ball (2015; 2018), Granger causality analysis is implemented in this study by using a standard VAR model. As discussed, the Akaike Information Criterion (AIC) is used to select an optimal maximum lag length (k) for the variables in the VAR. This lag length is then lengthened, if necessary, to ensure that firstly the estimated model is dynamically stable (i.e. all of the inverted roots of the characteristic equation lie inside the unit circle), and secondly, the errors of the equations are serially independent.

Granger causality results in this study are reported only if the VAR models meet the criteria in the preceding paragraph.

This study requires testing for Granger causality between the levels of some of the data series. In this case, the Granger causality testing procedure must be modified to allow for the differences in the orders of integration of the data series. Here, for each VAR model, the maximum lag length (k) is determined, but then one additional lagged value of each of the two variables is included in each equation of the VAR.

However, the Wald test for Granger non-causality is applied only to the coefficients of the original k lags. Toda and Yamamoto (1995) show that this modified Wald test statistic will still have an asymptotic distribution that is chi-square, even though the series is non-stationary, and the Granger causality test will be reliable.

**2.5.2.1 Lag-length selection for Granger causality testing**

In this paper, over the several dozen VARs run, the AIC sometimes suggests a large number of lags (numbering 40 and more) for a particular VAR (see Table 9).

The following points are made in this connection.

Firstly, Clarke and Mirza (2006) note that the choice of lag length in the VAR is important in order to avoid spurious causality or spurious absence of causality. They identify that the AIC has a positive probability of overestimating the true lag order, but suggest that overestimation of the lag order may be preferable, due to the problems that can occur with incorrect inferences in an underspecified model. Clarke and Mirza (2006) considered up to 10 lags of quarterly data, (a time span of 40 lags for monthly data) when choosing the systems lag order for a VAR model. In their selected models, the AIC criterion led to six and eight quarterly lags, respectively, equivalent to 24 and 32 monthly lags.

Turning to climate series, according to Mudelsee (2010) and Wilks (2011) autocorrelation in the atmospheric sciences (also called dependence, persistence or 'memory') is characteristic of many types of climatic fluctuations.

Further, what is termed long-range dependence can be seen in climate series, including atmospheric temperature series. For example, Franzke (2012) investigating trends in four temperature time series – Central England Temperature (CET), Stockholm, Faraday-Vernadsky, and Alert – found evidence of long-range dependence. Having established this, Franzke states that "Predictions of a long-range-dependent process need the knowledge of the whole past to predict the next state."

Further on climate series, using annual data, Triacca (2005) chose a maximum lag order of 10 (equivalent to 60 months), stating that it was important for the maximum order for the VAR to be high enough that high-order VAR specifications are given a reasonable chance of getting selected, if they happen to be appropriate.

Finally, the statistical package used in this study, the widely used EViews (IHS EViews, 2017) contains extensive error detection to prevent inappropriate model set-up in the statistical sense. The relevant error message reads: "Insufficient number of observations for specified maximum lag length." When this message appears the VAR cannot be run. All VARs reported were run and therefore by definition contained sufficient observations for the specified maximum lag length.

In conclusion, while some of our analyses led to lowest AIC results with large lag orders, with the above background, we consider these lag lengths to be legitimate for this study.

**2.5.2.2 Reporting of Granger causality results**

In the Granger causality results reported in each case, the relevant EViews output from the VAR model entitled 'VAR Granger causality/block Exogeneity' is used and the Wald Statistic *(p*-value) is reported. The extended EViews output for each Granger causality analysis is available in the Supplementary Information that accompanies this paper.

**2.6 Data**

For global surface temperature, we use the Hadley Centre–Climate Research Unit combined Landsat and SST monthly surface temperature series (HadCRUT) version 4.6.0.0 (Morice et al., 2012). In the paper, this series is abbreviated to 'At_temp'.

For atmospheric $CO_2$ data from 1958 to the present, we use monthly data from the $CO_2$ series produced by the US Department of Commerce National Oceanic and Atmospheric Administration Earth System Research Laboratory Global Monitoring Division Mauna Loa, Hawaii (Keeling et al., 2009). In the paper, this series is termed '$CO_2$'.

To represent global surface temperature models, a monthly data series projected from a business-as-usual global climate model (GCM) for global surface temperature was used. We used the CMIP5, RCP4.5 scenario model (Taylor et al., 2012) run for the IPCC fifth assessment report (IPCC, 2013)

Normalized Difference Vegetation Index (NDVI) monthly data from 1980 to 2006 come from the Global Inventory Modeling and Mapping Studies (GIMMS) dataset (Tucker et al., 2005); NDVI monthly data from 2006 to 2013 were provided by the Institute of Surveying, Remote Sensing and Land Information, University of Natural Resources and Life Sciences, Vienna. Being from the same disaggregated data set, the two series are readily merged for use (for method see Leggett and Ball, 2015) and the merged series is termed 'NDVI'.

The highest frequency data available for ocean heat content is reported quarterly. Quarterly data for ocean heat content data from the surface to 700 metres is from the NOAA National Oceanographic Data Center (NODC) (Levitus et al., 2012) for the period 1955 to the present. According to Climate Change Synthesis Report AR4 (IPCC, 2007), two-thirds of energy stored in the ocean is absorbed between the surface and a depth of 700 metres. The ocean heat from surface to 700 metres series as used in the paper is termed 'ocean heat uptake'.

Volcanic aerosol data is from the Stratospheric Aerosol Optical Thickness monthly series produced by the National Aeronautic and Space Administration Goddard Institute for Space Studies (Sato et al., 1993). It has previously been shown that increasing volcanic aerosols correlate with reduced temperature (Yan et al., 2016). In radiative forcing terms, the forcing from volcanic aerosols is approximately 27 times the optical thickness (Stern and Kaufmann, 2014; Pasini et al., 2017). Hence, the volcanic aerosol series used in the paper is sign-reversed and is termed 'reverse volcanic aerosols'.

Two cloud cover data sets are used – ocean and land. The monthly cloud cover data set used for ocean cloud is the Cloudiness Monthly Mean at Surface ICOADS v2.5 cloud cover (Freeman et al., 2016). The monthly land cover data set used is the CRUTS 4.03 (land) series (Harris et al., 2014).

Wind speed data comes from the monthly global ICOADS v2.5 wind speed (Variable wind speed – Scalar Wind Monthly Mean at Surface – in m/s) (Freeman et al., 2016).

Southern Oscillation Index (SOI) monthly data (Troup, 1965) are from the Science Delivery Division of the Department of Science, Information Technology, Innovation and the Arts (DSITIA), Queensland, Australia.

Outgoing longwave radiation (OLR) monthly data was accessed from Climate Explorer (Lee et al., 2007).

## 3. Results

This section is structured as follows. Firstly, we scan for candidates for roles in the control system as considered from a number of perspectives: either as physical elements, as outcomes, or as control terms. We then reassess the number of control terms in the control system model. Following this, we conduct Granger causality analysis on the selected time series under the terms of the control system hypothesis.

Finally, we scan the results to select chains of elements that pass the tests, which therefore suggest that they comprise the physical control system.

### 3.1 Candidate elements of the control system

In this section, as set out in the hypothesis (Section 1.2), we develop a list of putative candidate elements for a global atmospheric surface temperature control system.

The state of the atmosphere involves the major phenomena of temperature, humidity, precipitation, air pressure, wind speed, and cloud cover (Schneider et al., 2011).

These constitute potential candidates for the entities of the hypothesis: leading element, controller term, candidate actuator and outcome.

### 3.1.1 Outcomes of the control system

We address two terms under this heading: final outcome and penultimate outcome.

#### 3.1.1.1 Final outcome of the control system

The final outcome of the global atmospheric temperature control system is, by definition, global atmospheric temperature.

### 3.1.1.2 Candidate penultimate outcomes of the control system

An overall scan of atmospheric series leads us to nominate a further outcome class: the penultimate outcome. Penultimate outcomes of the control system are temperature-related and contribute to the final temperature outcome of the control system.

One candidate penultimate outcome of the control system selected for assessment is outgoing longwave radiation (OLR). This creates a cooling control by emitting infrared radiation from the surface of the earth (IPCC, 2013). It is part of terrestrial radiation, which is radiation emitted by the Earth's surface, the atmosphere and the clouds. As it may be affected by the degree of global cloud cover, it is considered a penultimate outcome contributing to the final outcome.

A second candidate penultimate outcome would arise if the control system enhanced heat uptake from the atmosphere by the ocean, thereby cooling the atmosphere (IPCC, 2013).

### 3.1.2 Candidate actuators of the control system

As this is a permutational study, we seek to minimise the number of permutations generated, by reducing the number of variables involved in the study wherever possible and reasonable.

We look for candidate actuators from major atmospheric phenomena previously outlined (that is, temperature, humidity, precipitation, air pressure, wind speed, and cloud cover (Schneider et al., 2011)).

Of these phenomena, it can be shown that there are close correlations between humidity, cloud cover and precipitation (for example, Richards and Arkin,1981; Walcek, 1994). For this study, we choose cloud cover to represent these phenomena.

We therefore selected atmospheric pressure, wind speed and cloud cover as candidate actuators for this study.

We note that cloud cover is assessed not only for a relationship from actuator to penultimate outcome but also directly from actuator to final outcome.

### 3.1.2.1 Atmospheric pressure

The trend in atmospheric pressure is commonly indicated by the Southern Oscillation Index (SOI) (IPCC, 2013), which is used in this study. The SOI is one of the two components of the El Niño–Southern Oscillation (ENSO).

### 3.1.2.2 Wind Speed

A large body of literature supports the inference that wind stress (force per unit area exerted by the wind on the ocean) directly controls the mass fluxes in the top several hundred meters of the ocean (Wunsch, 2002). Wunsch notes that the ocean is both heated and cooled within about 100 m of the surface.

Sarmiento and Gruber (2006) also state that winds are the main source of energy for ocean circulation, producing wind-driven ocean currents. A theory exists that biosphere-enhanced evaporation and condensation plays a general and dominant role in atmospheric dynamics (for overview see Sheil, 2018).

We thus have a potential pathway for the control system from the biosphere to the ocean.

Given this background, wind is considered to be a candidate actuator for moving heat from the atmosphere to the ocean.

### 3.1.2.3 Cloud cover

According to IPCC (2013), clouds can act to warm the Earth by trapping outgoing longwave infrared radiative flux at the top of the atmosphere. Lower level clouds can also cool the Earth by reflecting shortwave solar radiative flux back to space. The cloud radiative effect on the Earth's radiation budget can be inferred from satellite data by comparing upwelling radiation in cloudy and non-cloudy regions. This comparison shows that cloud conditions exert a global and annual shortwave cloud radiative effect of approximately -50 $W/m^2$ and a mean longwave cloud radiative effect of approximate 30 $W/m^2$. The net global mean cloud radiative effect is approximately -20 $W/m^2$ implying a strong net cooling effect of clouds on the current climate. A cooling effect with Granger causal characteristics of cloud albedo on global surface temperature is sought in this study.

As well, clouds modify the general circulation and hydrologic cycle through their interactions with the atmosphere, ocean, and land (IPCC, 2013). Hence cooling effects of clouds might also be expected on longwave radiation, and this is explored for by assessing Granger causality between clouds and outgoing longwave radiation.

Both land and ocean cloud cover time series are used in the analysis.

### 3.1.2.4 Candidate control terms

The control terms initially used in this analysis were the PID terms from Leggett and Ball (2020), termed here P_CONTR, I_CONTR and D_CONTR.

#### 3.1.2.4.1 Number of control terms in the control system

In Leggett and Ball (2020) we provided evidence that the atmospheric surface temperature control system had Proportional-Integral-Derivative (PID), that is, 3-term, control system characteristics.

It is well known that SOI (ENSO) and $CO_2$ variability are linked (Bacastow, 1976; for overview see Imbers et al., 2013.) In particular, Leggett and Ball (2015) showed that SOI correlated with second-difference $CO_2$, and this is the relationship that we explore further here. The relationship is shown in Figure 1, and the relevant ARDL analysis is provided in Tables 2 and 3.

Figure 1. Second-difference $CO_2$ (blue curve) and Reverse Southern Oscillation Index (red curve); monthly data

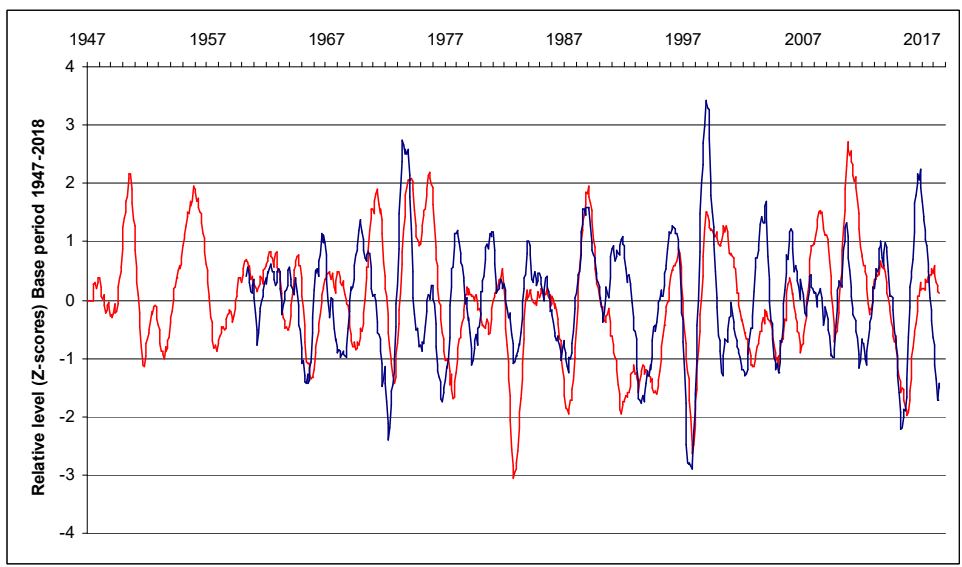

Table 2. Eviews ARDL estimation output for period May 1960 to July 2018 for SOI as a function of second-difference $CO_2$: overall statistics for short-run model dynamic relationship

| Model dependent variable | Number of models evaluated | Selected model | Autocorrelation out to 12 lags | Adjusted R-squared | F-statistic | p-value |
|---|---|---|---|---|---|---|
| SOI_CONTR | 156 | ARDL (4,2) | Nil | 0.987121 | 7610.544 | < 10-100 |

Table 3. Eviews ARDL estimation output for period May 1960 to July 2018 for SOI as a function of second-difference $CO_2$: statistics for each dependent variable in short-run model dynamic relationship

| Model independent variables | Coefficient | Std. Error | t-Statistic | Prob.* |
|---|---|---|---|---|
| SOI_CONTR(-1) | 1.337157 | 0.037865 | 35.31336 | 0.00E+00 |
| SOI_CONTR(-2) | -0.1111 | 0.06333 | -1.75436 | 7.98E-02 |
| SOI_CONTR(-3) | -0.16888 | 0.062944 | -2.683005 | 7.50E-03 |
| SOI_CONTR(-4) | -0.08494 | 0.037848 | -2.244243 | 0.0251 |
| DD_CO2_CONTR | 0.082316 | 0.027114 | 3.035941 | 0.0025 |
| DD_CO2_CONTR(-1) | -0.01275 | 0.047181 | -0.270272 | 0.787 |
| DD_CO2_CONTR(-2) | -0.05543 | 0.02721 | -2.037153 | 0.042 |
| C | -0.00136 | 0.004421 | -0.308568 | 0.7577 |

Tables 2 and 3 show that SOI is statistically significantly correlated with second-difference $CO_2$ (DD_$CO_2$). This would suggest that the control system may not be simply a PID system, but may include an extra control term reflecting second difference (that is, a PID+DD system). In man-made engineering, such systems exist and can perform better than PID systems alone (Sahib, 2015; Raju et al., 2016).

We investigate this possibility for the atmospheric temperature control system as follows. ARDL analysis is carried out for both PID and PID+DD combinations. If the PID+DD combination displays a higher adjusted R-squared value and an improved Information Criterion, then this provides evidence for the control system being of PID+DD type.

The results of this analysis are presented in Tables 4 to 7.

Table 4. Eviews ARDL estimation output for period May 1960 to July 2018 for atmospheric surface temperature as a function of putative control system P_CONTR, I_CONTR and D_CONTR terms: overall statistics for short-run model dynamic relationship

| Model dependent variable | Number of models evaluated | Selected model (SIC) | Autocorrelation out to 36 lags | Adjusted R-squared | F-statistic | p-value | Akaike Information Criterion |
|---|---|---|---|---|---|---|---|
| At_temp | 3,584 | ARDL(4, 0, 0, 5) | Nil | 0.902242 | 539.3778 | 0.00E+00 | 0.525009 |

Table 5. Eviews ARDL estimation output for period May 1960 to July 2018 for temperature as a function of putative control system P_CONTR, I_CONTR and D_CONTR terms: statistics for long-run model independent variables

| Model independent variables | Coefficient | Std. Error | t-Statistic | Prob.* |
|---|---|---|---|---|

| | | | - | |
|---|---|---|---|---|
| P_CO2_CONTR | -1.687791 | 0.596673 | 2.828668 | 0.0048 |
| I_CO2_CONTR | 1.013032 | 0.598024 | 1.693965 | 0.0907 |
| D_CO2_CONTR | -0.290399 | 0.087033 | -3.336661 | 0.0009 |
| C | 0.181349 | 0.051687 | 3.508626 | 0.0005 |

Table 6. Eviews ARDL estimation output for period May 1960 to July 2018 for atmospheric surface temperature as a function of putative control system P_CONTR, I_CONTR, D_CONTR and DD_CONTR terms: overall statistics for short-run model dynamic relationship

| Model dependent variable | Number of models evaluated | Selected model | Autocorrelation out to 36 lags | Adjusted R-squared | F-statistic | p-value | Akaike Information Criterion |
|---|---|---|---|---|---|---|---|
| At_temp | 28,672 | ARDL(4, 1, 0, 2, 0) | Nil | 0.903408 | 594.4807 | 0.00E+00 | 0.507034 |

Table 7. Eviews ARDL estimation output for period May 1960 to July 2018 for temperature as a function of putative control system P_CONTR, I_CONTR, D_CONTR and DD_CONTR terms: statistics for long-run model independent variables

| Model independent variables | Coefficient | Std. Error | t-Statistic | Prob.* |
|---|---|---|---|---|
| P_CO2_CONTR | -1.9408 | 0.6487 | -2.99183 | 0.0029 |
| I_CO2_CONTR | 1.4016 | 0.654659 | 2.140961 | 0.0326 |
| D_CO2_CONTR | -0.519331 | 0.104663 | -4.961912 | 8.81E-07 |
| DD_CO2_CONTR | -0.25982 | 0.098192 | -2.646033 | 0.0083 |
| C | 0.370506 | 0.103063 | 3.594928 | 0.0003 |

Tables 4 to 7 show that the DD term is significant in the four-term control type model and that this model displays an improved adjusted R-squared value and Akaike Information Criterion value. Note that the I term also displays a higher R-squared value (that is, displays a better fit) in the four-term model than in the three-term model.

We conclude that the above is statistically significant evidence that the atmospheric temperature control system is of four-term PID+DD type.

This presents a practical consequence for the remainder of this study – in the connectivity analysis, we assess the DD term alongside the P, I and D terms.

**3.2 Candidate leading element**

The most prominent global-level control system hypothesis proposes that the control system is biotic in origin (Lovelock and Margulis, 1974). Here, Lovelock and Margulis referred to "…the notion of the biosphere as an active adaptive control system able to maintain the Earth in homeostasis..." They referred to this as the 'Gaia' hypothesis.

In Leggett and Ball (2020), we noted that the control system for which we provided evidence carries out a task similar to that which Gaia is hypothesised to perform. However that study did not go so far as to identify a physical mechanism.

The biosphere is hence a candidate in this study for being the lead element of the control system. It can be considered to be the physical entity in which the controller resides – the controller outputs would clearly not constitute all of the outputs of the biosphere.

An indicator of the action of the biosphere is through measurement of its extent, and the amount of its photosynthesis. There is no single time series available for global photosynthesis. Photosynthesis takes place on land and in the oceans.

In considering the separate land and ocean domains, on the one hand the extent of photosynthesising *biomass* is dominated by land with 450 Gt C compared with the oceans at 2 Gt C (Bar-On et al., 2018). On the other hand, the relative contributions of land and oceans in *photosynthesis* are much closer together, with terrestrial sources representing 53.8% and oceanic sources 46.2% (Field et al., 1998).

Given the overall dominant position of terrestrial photosynthesis in both of these two measures, any biosphere control system effects are likely to be seen from the terrestrial photosynthesis trend. The global terrestrial photosynthesis trend is therefore considered suitably representative of the biosphere for use in this study.

An accepted measure of global terrestrial photosynthesis is through satellite measurements of vegetation reflectance, such as the Normalized Difference Vegetation Index (NDVI). We use the trend in NDVI to indicate that of the global biosphere in this study.

We now address the manner in which the biosphere might specifically affect climate, and in particular from the perspective of the candidate control system actuators of wind speed and cloud cover.

With respect to wind speed, Sheil (2018) reviews evidence that biosphere-enhanced evaporation and condensation play a general and dominant role in the generation of atmospheric motion – global circulation patterns, including winds.

Hence we have a potential causality pathway from the NDVI candidate leading element to the wind speed candidate actuator.

Turning to cloud cover, according to Gordon et al. (2016) most cloud droplets need tiny airborne particles to act as 'seeds' for their formation and growth. More particles in the atmosphere lead to more reflective clouds and a cooler climate. Trees produce a large fraction of cloud 'seeds' over the cleanest forested parts of the world. Such cloud seeds are particles generated by trees as a result of their emission of terpene gases (Gordon et al. (2016).

Hence we have a further potential causality pathway from the NDVI candidate leading element, via production of terpenes, to the cloud cover candidate actuator.

### 3.3 Developing a possible structure for the control system

After identifying candidate elements and then sorting them by type, we arrive at a hypothesised four-step control sequence: leading element to each controller term; each controller term to each actuator; each actuator to each penultimate outcome; and each penultimate outcome to final outcome. With five candidate actuators, two penultimate outcomes and one final outcome, the permutations of the foregoing provide the hypothesised control system shown in Table 8.

Table 8. Candidate control system elements in the combinations to be assessed for connectivity by Granger causality analysis

| Control system step 1 | | Control system step 2 | | Control system step 3 | | Control system step 4 | |
|---|---|---|---|---|---|---|---|
| Leading element to | Controller term | Controller term to | Actuator | Actuator to | Penultimate outcome | Penultimate outcome to | Final outcome |
| NDVI_CONTR | P_CO2_CONTR | P_CO2_CONTR | Wspeed_CONTR | Wspeed_FROM_2002_CONTR | Ocean heat uptake_REVERSE | Ocean heat uptake_REVERSE | At_temp |
| NDVI_CONTR | I_CO2_CONTR | P_CO2_CONTR | Cloud_Ocean_CONTR | Wspeed_CONTR | OLR_REVERSE | OLR_REVERSE | At_temp |
| NDVI_CONTR | D_CO2_CONTR | P_CO2_CONTR | Cloud_Land_CONTR | Cloud_Ocean_CONTR | Ocean heat uptake_REVERSE | | |
| NDVI_CONTR | DD_CO2_CONTR | P_CO2_CONTR | SOI_CONTR | Cloud_Ocean_CONTR | OLR_REVERSE | | |
| | | P_CO2_CONTR | PNA_CONTR | Cloud_Land_CONTR | Ocean heat uptake_REVERSE | | |
| | | I_CO2_CONTR | Wspeed_CONTR | Cloud_Land_CONTR | OLR_REVERSE | | |
| | | I_CO2_CONTR | Cloud_Ocean_CONTR | SOI_CONTR | Ocean heat uptake_REVERSE | | |
| | | I_CO2_CONTR | Cloud_Land_CONTR | SOI_CONTR | OLR_REVERSE | | |
| | | I_CO2_CONTR | SOI_CONTR | PNA_CONTR | Ocean heat uptake_REVERSE | | |
| | | I_CO2_CONTR | PNA_CONTR | PNA_CONTR | OLR_REVERSE | | |
| | | D_CO2_CONTR | Wspeed_CONTR | | | | |
| | | D_CO2_CONTR | Cloud_Ocean_CONTR | | | | |

| | |
|---|---|
| D_CO2_CONTR | Cloud_Land_CONTR |
| D_CO2_CONTR | SOI_CONTR |
| D_CO2_CONTR | PNA_CONTR |
| DD_CO2_CONTR | Wspeed_CONTR |
| DD_CO2_CONTR | Cloud_Ocean_CONTR |
| DD_CO2_CONTR | Cloud_Land_CONTR |
| DD_CO2_CONTR | SOI_CONTR |
| DD_CO2_CONTR | PNA_CONTR |

All of the variable pairs are tested for Granger causality. This process will provide empirical evidence about the relationship between each pair of elements in the hypothesis.

A flow of one-way Granger causality across each of the four control system stages in at least one of the four P, I, D, or DD control type 'channels' would provide evidence of at least some of the elements of a physical control system..

It is noted that a thorough search for Granger causality is carried out. If it exists over only part of the time span of the series for a pair of elements, then it is included.

The next part of the analysis is to assess the information flow pattern found, and identify (i) any A to B relationships that are non Granger-causal, and (ii) any B to A Granger-causal relationships. Any such relationship even in a single step in a given candidate control system channel suggests on the face of it that the channel does not exist.

**3.4 Potential scale of influence of each candidate actuator**

To enable consideration of the nature of their potential influence, we visually depict each of the candidate control system elements. Candidate process (controller) elements are shown in Figure 2 and candidate physical (actuator) elements are shown in Figure 3. The data series used are from 1901 to 2018 and are Z-scored from 1901 to 1987 in order to show any points of departure from prior trends in the later period (1988 to 2018). Figures 2 and 3 show the candidate control system elements compared against (i) the expected temperature from the output of a mid-range scenario model (CMIP5, RCP4.5 scenario) (Taylor et al., 2012) run for the IPCC fifth assessment report (IPCC, 2013) and with (ii) observed temperature.

Figure 2. Trends in expected temperature from the output of a mid-range IPCC scenario model (CMIP5, RCP4.5 scenario) (dark blue curve), observed temperature (blue curve), and control system candidate process elements: P_$CO_2$_CONTR (black curve); I_$CO_2$_CONTR (turquoise curve); D_$CO_2$_CONTR (brown curve); and DD_$CO_2$_CONTR (red curve); annual data, 1901 to 2018; all data Z-scored, base period 1901-1987

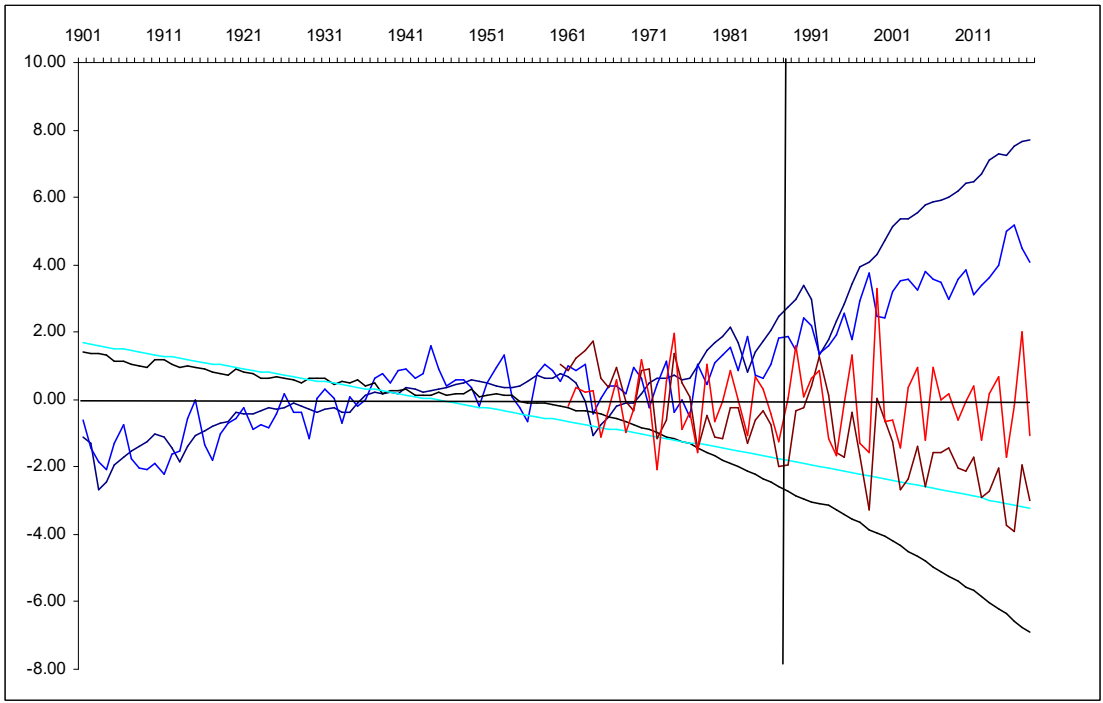

Figure 3. Trends in expected temperature from the output of a mid-range IPCC scenario model (CMIP5, RCP4.5 scenario)(dark blue curve), observed temperature (blue curve), and control system candidate physical elements: Ocean heat (black curve); NDVI (green curve); Wind speed (purple curve); OLR (brown curve); Ocean cloud (black dashed curve); Land cloud (turquoise curve); and SOI (red curve); annual data, 1901 to 2018; all data Z-scored, base period 1901-1987

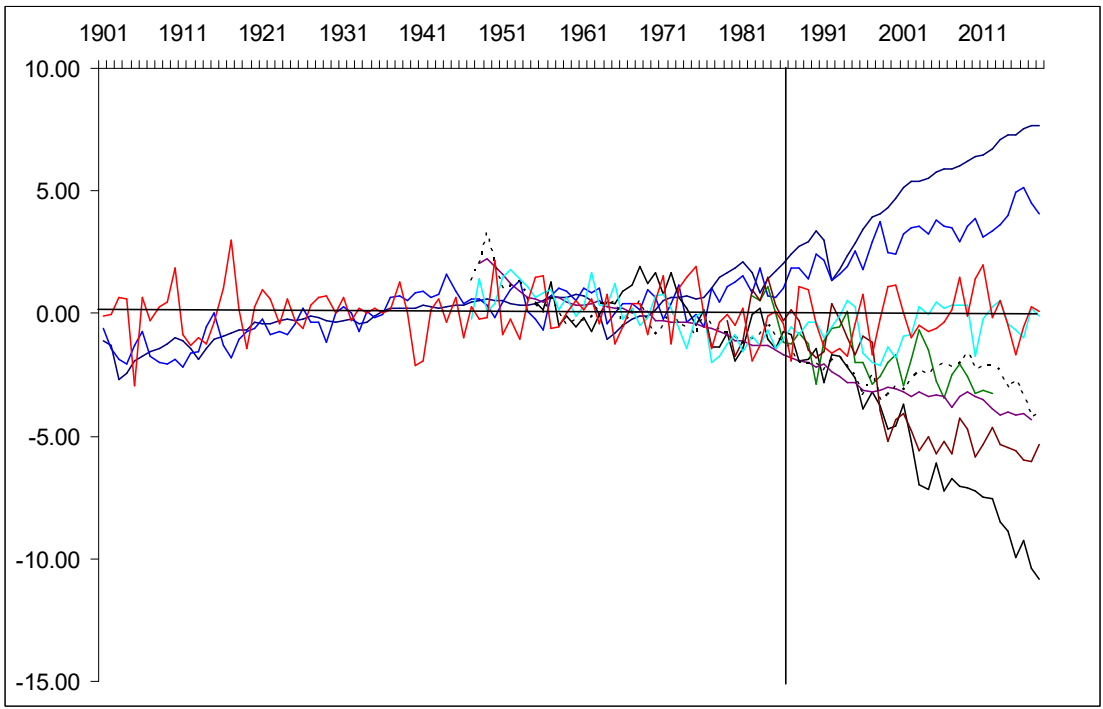

Figures 2 and 3 show that the trends of the potential control system elements fall into three main classes in terms of degree of potential temperature reduction, based on their deviation from the level in the base period (1901 to 1987). These are: from slight or none – DD_$CO_2$_CONTR, SOI, PNA and land cloud; to moderate – D_$CO_2$_CONTR, OLR_REVERSE, NDVI, wind speed and ocean cloud; to largest – P_$CO_2$_CONTR, I_$CO_2$_CONTR, and ocean heat uptake. These trends will be further referred to after the results of causality analysis are presented in Table 9.

### 3.5 Results of assessments of Granger causality for candidate control elements

Results of Granger causality testing between candidate control elements are presented first for link-by-link causal relationships, and then for long-range causal relationships.

Table 9 presents the results of the Granger causality assessments at link-by-link level.

### 3.5.1 Basis of Granger causality testing for link-by-link causal relationships between candidate control elements

In general, the results in Table 9 are from bivariate Granger causality assessments, with an exception in four cases (noted in Table 4 of the Supplementary Information) involving trivariate analysis. The four cases involving trivariate analysis will be addressed before discussing the overall results.

### 3.5.1.1 Trivariate Granger causality analysis: NDVI and control system process terms

In bivariate Granger causality analyses for NDVI and the control system process terms (P_$CO_2$_CONTR, I_$CO_2$_CONTR, D_$CO_2$_CONTR and DD_$CO_2$_CONTR) we explored the lag space extensively but could not find causality.

This situation was explored further by conducting Granger causality analysis using a third, auxiliary, variable in the causality analyses (for example, see Triacca et al. (2013).

A possible candidate for the auxiliary variable was volcanic aerosols (abbreviated here to 'Volc'). The relationship between NDVI and Volc is shown in Figure 4.

Figure 4. Trends in NDVI (red line) and volcanic aerosols (sign reversed – blue line). Also shown is the linear trend for volcanic aerosols (black line); annual data, 1947 to 2018; all data *Z*-scored

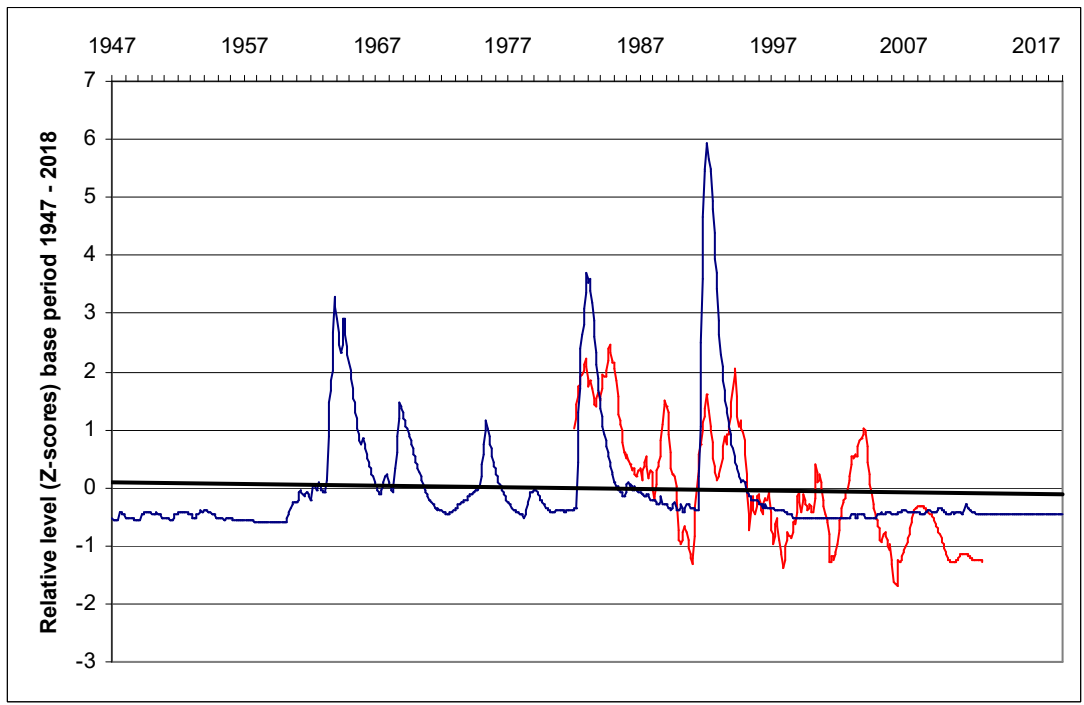

Figure 4 shows that, while displaying an essentially neutral long-term trend, short-term variability in the volcanic aerosol series correlates with NDVI. As a result, volcanic aerosols might be appropriate as an auxiliary variable.

Therefore, trivariate Granger causality analyses were conducted for NDVI with Volc and each of the control system process terms P_$CO_2$_CONTR, I_$CO_2$_CONTR, D_$CO_2$_CONTR and DD_$CO_2$_CONTR. This approach found causality for the $CO_2$ variables, with the results being reported in Table 9.

For clarity, following Triacca et al. (2013), only results for NDVI and the $CO_2$ terms are reported in Table 9. Full results for the trivariate analysis including Volc are provided in the Supplementary Information.

### 3.5.2 Results of assessments of Granger causality for candidate control elements at link-by-link level

Table 9 presents the results of all Granger causality testing to be scanned for full chains displaying one-way causality in line with the control system hypothesis.

Information is provided in the table on the degree of smoothing of each series (when used), lag length selected, autocorrelation taken into account, and whether the Toda-Yamamoto procedure was used. (More extensive information on Granger causality attributes for each assessment is provided in the Supplementary Information.)

In the table, the single best evidence for causality from a particular Granger causality assessment is shown. Hence the assessment in the table may be from either monthly or quarterly data. The type of data used for each assessment is specified in the Supplementary Information.

Bold figures in Table 9 represent a statistically significant relationship (that is, a Granger-causal relationship with a probability of being due to chance of 0.05 or less). A non-significant relationship is presented in italics. Therefore, a relationship in line with the hypothesis – that is, showing one-way Granger causality in the A to B direction – will be displayed with bold numbering in the A to B column and italics in the B to A column.

Table 9. Results of assessments of Granger causality for candidate control elements. Abbreviations not otherwise explained in text: T-Y – Toda-Yamamoto; AIC – Akaike information criterion. Data series used are monthly except where marked with the suffix 'quar'. The Granger causality assessment for Wind speed to Ocean heat uptake is based on data from 2002 onwards (see text)

| Row | Control system step 1 | | | | | | Control system step 2 | | | | | | | Control system step 3 | | | | | | | Control system step 4 | | | | | | | |
|---|---|---|---|---|---|---|---|---|---|---|---|---|---|---|---|---|---|---|---|---|---|---|---|---|---|---|---|---|
| | Biosphere leading element | Controller term | | | | Probability that relationship not Granger causal | Controller term | Actuator | | | | Probability that relationship not Granger causal | | Actuator | Penultimate outcome | | | | Probability that relationship not Granger causal | | Actuatir or Penultimate outcome | Final outcome | | | | Probability that relationship not Granger causal | |
| | Variable A | Variable B | Lags (AIC) | T-Y test used | No autocor-relation out to n lags | Causality A to B | Causality B to A | Variable A | Variable B | Lags (AIC) | T-Y test used | No autocor-relation out to n lags | Causality A to B | Causality B to A | Variable A | Variable B | Lags (AIC) | T-Y test used | No autocor-relation out to n lags | Causality A to B | Causality B to A | Variable A | Variable B | Lags (AIC) | T-Y test used | No autocor-relation out to n lags | Causality A to B | Causality B to A |
| 1 | NDVI_CONTR | P_CO2_CONTR | 74 | N | 24 | 1.58E-05 | 0.4123 | P_CO2_CONTR | Wspeed_CONTR | 41 | Y | 12 | 0.0268 | 0.1847 | WSPEED_FROM_2002_CONTR_Quar | Ocean heat uptake_REVERSE_quar | 8 | Y | 8 | 0.0029 | 0.1377 | Cloud_Ocean_CONTR | At_temp | 14 | Y | 11 | 0.0145 | 0.3109 |
| 2 | NDVI_CONTR | I_CO2_CONTR | 74 | N | 9 | 0.0069 | 0.2394 | P_CO2_CONTR | Cloud_Ocean_CONTR | 61 | Y | 7 | 0.0766 | 0.6547 | Wspeed_CONTR | OLR_REVERSE | 40 | Y | 12 | 0.0055 | 0.1077 | Cloud_Land_CONTR_quar | At_temp | 12 | N | 13 | 0.0034 | 0.2047 |
| 3 | NDVI_CONTR | D_CO2_CONTR | 49 | N | 13 | 0.0124 | 0.3303 | P_CO2_CONTR | Cloud_Land_CONTR | 54 | Y | 12 | 0.0143 | 0.1987 | Cloud_Ocean_CONTR_quar | Ocean heat uptake_REVERSE_quar | 27 | Y | 12 | 0.7044 | 0.626 | Ocean heat uptake_REVERSE_quar | At_temp_quar | 5 | Y | 6 | 0.0017 | 0.0702 |
| 4 | NDVI_CONTR | DD_CO2_CONTR | 46 | N | 12 | 0.023 | 0.355 | P_CO2_CONTR | SOI_CONTR | 67 | N | 20 | 0.0009 | 0.0537 | Cloud_Ocean_CONTR | OLR_REVERSE | 55 | Y | 20 | 0.0157 | 0.1029 | OLR_REVERSE | At_temp | 28 | Y | 6 | 0.0045 | 0.2533 |
| 5 | | | | | | | | I_CO2_CONTR | Wspeed_CONTR | 27 | Y | 15 | 0.0141 | 0.2104 | Cloud_Land_CONTR_quar | Ocean heat uptake_REVERSE_quar | 40 | Y | 20 | 0.4002 | 0.5482 | | | | | | | |
| 6 | | | | | | | | I_CO2_CONTR | Cloud_Ocean_CONTR | 15 | N | 7 | 0.0222 | 0.0148 | Cloud_Land_CONTR | OLR_REVERSE | 27 | Y | 12 | 0.0006 | 0.1338 | | | | | | | |
| 7 | | | | | | | | I_CO2_CONTR | Cloud_Land_CONTR | 27 | N | 11 | 0.0881 | 0.01 | SOI_CONTR_quar | Ocean heat uptake_REVERSE_quar | 41 | Y | 20 | 0.1283 | 0.4654 | | | | | | | |
| 8 | | | | | | | | I_CO2_CONTR | SOI_CONTR | 41 | N | 16 | 0.0196 | 0.0076 | SOI_CONTR | OLR_REVERSE | 30 | Y | 20 | 0.1713 | 0.0503 | | | | | | | |
| 9 | | | | | | | | D_CO2_CONTR | Wspeed_CONTR | 41 | Y | 9 | 0.054 | 0.1164 | | | | | | | | | | | | | | |
| 10 | | | | | | | | D_CO2_CONTR | Cloud_Ocean_CONTR | 80 | N | 12 | 0.0343 | 0.4853 | | | | | | | | | | | | | | |
| 11 | | | | | | | | D_CO2_CONTR | Cloud_Land_CONTR | 68 | N | 12 | 0.0474 | 0.8088 | | | | | | | | | | | | | | |
| 12 | | | | | | | | D_CO2_CONTR | SOI_CONTR | 29 | N | 10 | 0.004 | 0.0008 | | | | | | | | | | | | | | |
| 13 | | | | | | | | DD_CO2_CONTR | Wspeed_CONTR | 18 | Y | 4 | 0.0249 | 0.0079 | | | | | | | | | | | | | | |
| 14 | | | | | | | | DD_CO2_CONTR | Cloud_Ocean_CONTR | 67 | N | 9 | 0.0809 | 0.3354 | | | | | | | | | | | | | | |
| 15 | | | | | | | | DD_CO2_CONTR | Cloud_Land_CONTR | 81 | N | 12 | 0.0209 | 0.6945 | | | | | | | | | | | | | | |
| 16 | | | | | | | | DD_CO2_CONTR | SOI_CONTR | 80 | N | 5 | 0.001 | 0.07 | | | | | | | | | | | | | | |

We note at the outset that Table 9 shows that from trivariate analyses with volcanic aerosols as an auxiliary variable (see Section 3.5.1.1), the links from NDVI to the four $CO_2$ forms all show Granger causality in line with the hypothesis. This raises the question of whether or not volcanic aerosols must be considered to be part of the control system.

Volcanic aerosols act to decrease global surface temperature (IPCC, 2013). This would assist any control system to reach a set point that requires a cooling action. However we consider that volcanic aerosols are a factor external to the control system influencing the initial temperature it responds to, similar to the effect of clouds for the air conditioning system of a building, and therefore not a part of the control system itself.

In scanning Table 9, it can be seen that the control system step from ocean heat uptake to atmospheric temperature is shown to display one-way Granger causality in line with the control system hypothesis, but also that causality in the other direction (P = 0.07) would be significant were the 0.1 level of statistical significance to be chosen.

For this reason (as outlined in Section 1.5), the strength of causality $F_{XY}$ and $F_{YX}$ is calculated for each direction in this relationship. The results are $F_{Ocean\ heat\ to\ At.\ temp} = 0.1406$, and $F_{At.\ temp\ to\ Ocean\ heat} = 0.0565$. This result shows that causality is close to 2.5 times stronger in the direction of ocean heat uptake to atmospheric temperature than the reverse. We consider this as further evidence for effective one-way causality in line with the control system hypothesis.

Table 9 shows evidence for both actuator to final outcome and penultimate outcome to final outcome control system steps. These are, respectively: each of ocean cloud and land cloud to atmospheric temperature; and ocean heat uptake and outgoing longwave radiation to atmospheric temperature.

**3.5.3 End-to-end control system causal chains**

The results of the scan of Table 9 for causal chains in line with the control system hypothesis are assembled in Figure 5.

The figure shows the chains that display one-way causality across each of the four sequential control system stages from leading element to final outcome (note that for clarity, the 'Controllers' column displays the relevant controllers for each Actuator).

Figure 5. Evidence for the existence of a physical control system: end-to-end sequences of candidate control system elements that display one-way Granger causality at 0.05 probability level across each step of the sequence (derived from Table 9)

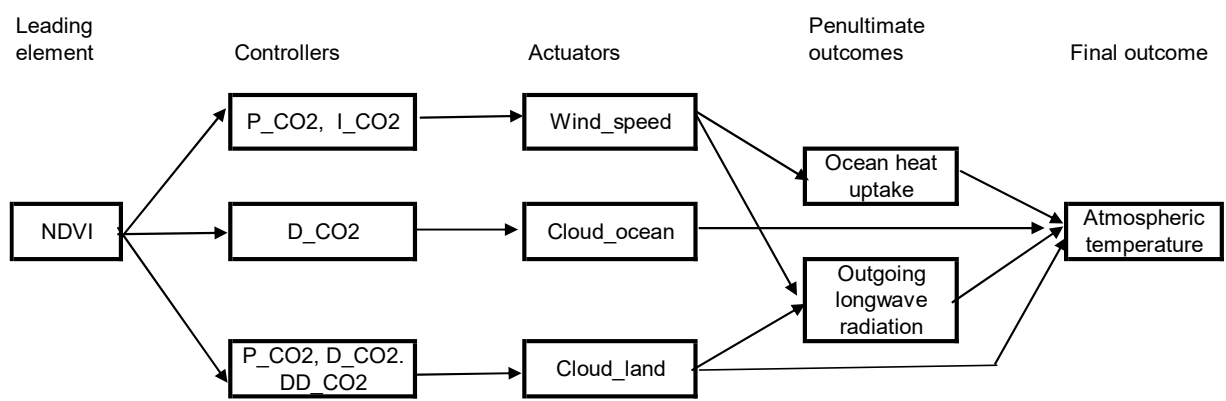

Figure 5 shows complete end-to-end one-way Granger causality pathways sufficient to meet the control system hypothesis.

### 3.5.4 Action of the leading element of the control system

Turning to the question of the leading element of the control system, the patterns of one-way Granger causality shown in Table 9 and summarised in Figure 5 show that the leading element is NDVI. Noting that, in this paper, NDVI represents the biosphere (Section 3.2), as previously outlined the most prominent global-level atmospheric control system that has been hypothesised also proposes that the control system is biotic in origin (Lovelock and Margulis, 1974). Here, Lovelock and Margulis referred to "…[t]he notion of the biosphere as an active adaptive control system able to maintain the Earth in homeostasis..."

They went on to write "...we are calling [this] the 'Gaia' hypothesis …"

In Table 9, then, NDVI stands for the Gaia candidate. The biosphere can be considered to be the physical entity in which the controller resides (the controller outputs being some of the outputs of the biosphere).

The pattern of results in Table 9, then, can be read to provide strong evidence of the existence of Gaia as proposed in the Gaia hypothesis.

The evidence of control system connectivity is from the leading element to each of the four control-type terms, through to three actuators and then via one actuator to the ocean heat uptake penultimate outcome and via all three actuators to the outgoing longwave radiation penultimate outcome. Both penultimate outcomes display one-way Granger causality to the final outcome. Hence there are multiple lines of evidence for the existence of the physical control system.

The likely physical actions to produce these control system results are described in the following.

Evidence was provided in Section 3.1.2.2 of wind activity induced by evaporation and condensation from the biosphere (Sheil, 2018) and in Section 3.2 of cloud cover being influenced by terpene production from the biosphere (Gordon et al., 2016).

The $CO_2$ controller terms (P_$CO_2$_CONTR, I_$CO_2$_CONTR, D_$CO_2$_CONTR and DD_$CO_2$_CONTR) represent the various ways that the biosphere manages its control system output of evaporation, condensation and terpenes in response to the disturbance of rising atmospheric $CO_2$. In addition to a direct linear increased output of evaporation, condensation and terpenes in response to the disturbance (P_$CO_2$_CONTR), it also demonstrates an increased output proportional to change of the disturbance (D_$CO_2$_CONTR), to the total disturbance (I_$CO_2$_CONTR), and also to the rate of change of the disturbance (DD_$CO_2$_CONTR).

Turning to actuators, it appears that global wind speed and global cloud cover (both land- and ocean-based) are likely the main actuators, with wind speed moving heat from the atmosphere to the ocean and cloud cover reflecting incoming solar radiation and allowing upwelling of terrestrial radiation to space.

These results, then, are evidence for the physical nature of the control system affecting global surface temperature.

### 3.5.5 Further observations on the relationship between Wind Speed and Ocean Heat Uptake

On the control relationship from wind speed to ocean heat uptake (an actuator to penultimate outcome relationship), while clear one-way Granger causality in line with the control system hypothesis was found (see Table 9) it was only over the period from 2002 to 2018, rather than the full time span of the data available to the study.

We therefore sought additional information to augment the above finding using the longer-range connection method (Section 1.4). This is done by seeking longer-range Granger causal relationships which span the direct link-to-link wind speed control to ocean heat uptake control relationship. Two relationships which span these links are assessed: controller term (D_CO2) to actuator (Wind speed control) and controller term (D_CO2) to penultimate outcome (Ocean heat uptake control). Results are presented in Table 10.

Table 10. Results of Granger causality tests for series pairs of Wind speed control and Ocean heat uptake control, D_CO2 control and Wind speed control and D_CO2 control and Ocean heat uptake control. Data series are monthly unless marked in the table as quarterly (Quar)

| Controller term | Actuator | Penultimate outcome | Lags (AIC) | T-Y test used | No auto-correlation out to n lags | Probability that relationship not Granger causal | |
|---|---|---|---|---|---|---|---|
| | | | | | | Causality A to B | Causality B to A |
| | WSPEED_from_2002_CONTR_Quar | OCEAN_CONTR_Quar | 8 | Y | 8 | **0.0029** | 0.1377 |
| D_CO2_CONTR from 1960 | WSPEED_CONTRO_from_1960 | | 41 | Y | 9 | **0.054** | 0.1164 |
| D_CO2_CONTR from 1960 | | OCEAN_CONTR_Quar-from 1960 | 6 | Y | 9 | **0.0206** | 0.0763 |

The first row in Table 10 reproduces the relevant result from Table 9. The second row is a further reproduction from Table 9 of the information showing one-way Granger causality between D_CO2_CONTR and the *input* to the relationship in the first row. The third row in the table shows the additional result of this analysis, that there is one-way Granger causality between D_CO2_CONTR and the *output* of the relationship in the first row.

It can be seen that the control system step from D_CO2 to Ocean heat uptake, while demonstrating one-way causality in the hypothesised direction, displays causality in the opposite direction, which while failing to reach significance at the 0.05 level would be significant at the 0.1 level.

For this reason (and as outlined in Section 1.5), the strength of causality $F_{XY}$ and $F_{YX}$ is calculated for each direction in this relationship. The results are $F_{D\_CO2\_Contr\ to\ Ocean\_Contr} = 0.1597$, and $F_{Ocean\_contr\ to\ D\_CO2\_Contr} = 0.0702$. This result shows that causality is about 2.3 times stronger in the direction of D_CO2_Contr to Ocean_Contr than the reverse. We consider this as further evidence for one-way causality in line with the control system hypothesis.

The pattern of one-way Granger causal results in Table 10 strengthens the evidence for the control relationship from wind speed to ocean heat uptake over the period from 2002 to 2018 (only part of the time span of the data available to the study) that was shown in Table 9. These results would appear to increase the likelihood that the relationship from wind speed to ocean heat uptake may exist for the full data time span available, and that the lack of a clear Granger causal result for the full time span may be due to as yet undetermined statistical or other factors.

**3.6 Long-range causality from candidate leading element to control system final outcome**

We now test for Granger causality from NDVI, shown in Table 9 to be the leading element of the control system, to atmospheric temperature, the final outcome of the control system.

By analogy with the electronic signal tracer test equipment used to troubleshoot radio and other electronic circuitry described in Section 1.4 (Bishop, 2007), this tests to confirm that the influence of the leading element appears in the outcome, confirming that the intervening elements are therefore connected as an unbroken chain.

Results are provided in Table 11.

Table 11. Results of Granger causality tests from leading element to control system final outcome

| Variable A | Variable B | Lags | No autocorr-elation out to n lags | Probability that relationship not Granger causal | |
|---|---|---|---|---|---|
| | | | | Causality A to B | Causality B to A |
| NDVI_CONTR | At_temp | 40 | 9 | **0.0113** | *0.1593* |

Table 11 shows that NDVI demonstrates clear statistically significant one-way Granger causality to the final outcome of the system, global surface temperature.

We suggest that this overall result serves to underline the detailed results shown in Table 9.

## 4. Discussion

By applying Granger causality analysis, this paper has provided empirical evidence relevant to a number of the aspects of the functioning of the control system affecting global atmospheric temperature: the physical nature of control system elements; the existence of full step-by-step one-way Granger causal control between the elements from leading element to outcome; the identification of four controller terms for the control system; a possible means for the physical connection of the biosphere, represented by NDVI, to the $CO_2$-related controller process terms; the actuators of the control system found and their relationship to the final outcome, global surface temperature, either directly or indirectly via penultimate outcomes; and evidence for the leading element of the control system being the global biosphere, and therefore Gaia.

The hypothesis put forward in this paper for the nature of the physical control system affecting global atmospheric temperature proposed a sequence of candidate element types making up the control system – the leading element, controllers, actuators, penultimate outcomes and final outcome. This paper provides evidence that one-way Granger causality is observed across each step of this proposed sequence of candidate element types. Hence these results are evidence for the physical existence of the control system, and of its nature.

We demonstrate that the control system shown by Leggett and Ball (2020) to contain three controller terms – proportional, integral and derivative (PID) – demonstrates an improvement in statistical fit when a fourth, second-derivative (approximated by second difference) or DD, controller term is included.

Clear, directional Granger causal or information flow connectivity is shown from NDVI to the four $CO_2$ controller terms and then to the actuators..

To support physical mechanisms, evidence was provided in Section 3.1.2.2 of wind activity induced by evaporation and condensation from the biosphere (Sheil, 2018), and in Section 3.2 of cloud cover being influenced by terpene production from the biosphere (Gordon et al., 2016).

In Section 3.5.4, we noted that the $CO_2$ controller terms (P_$CO_2$, I_$CO_2$, D_$CO_2$ and DD_$CO_2$) represent evidence for the various ways that the biosphere manages its control system output of evaporation, condensation and terpenes in response to the disturbance of rising atmospheric $CO_2$. In addition to a direct linear increased output of evaporation, condensation and terpenes in response to the disturbance (P_$CO_2$_CONTR), the evidence from this paper is for an increased output proportional to change of the disturbance (D_$CO_2$_CONTR), to the total disturbance (I_$CO_2$_CONTR), and also to the rate of change of the disturbance (DD_$CO_2$_CONTR).

Turning to actuators, we present evidence that that global wind and global cloud cover (both land- and ocean-based) are actuators. Cloud cover is shown to influence the final outcome, global surface temperature, directly by reflecting incoming solar radiation. Cloud cover and wind speed also influence the penultimate outcomes found, those of enhanced ocean heat uptake and enhanced outgoing longwave radiation, with wind speed moving heat from the atmosphere to the ocean and cloud cover showing causality to outgoing longwave radiation. These processes together lead to control system output to the final outcome, global atmospheric temperature.

We provide evidence that the leading element of the control system is the global biosphere, and show, by way of outlining the mechanisms of its action to control global atmospheric temperature, that this is evidence for the existence of Gaia as postulated by Lovelock and Margulis in 1974.

A notable feature of these results is that evidence for the activity of the control system is present across an extensive range of the major atmospheric series assessed in this paper.

Future research might be warranted into the possible existence of further processes of the control system involving, for example, seasonal timers and feed-forward (Åström and Murray, 2008). These might be further aspects enabling the precise application of coordinated control outputs worldwide into the global atmosphere.

# 5. References

Ahmad, N. and Du, L. 2017. Effects of energy production and $CO_2$ emissions on economic growth in Iran: ARDL approach. *Energy* **123**, 521–537. https://doi.org/10.1016/j.energy.2017.01.144

Allen, M.P. 1997. *Understanding regression analysis*. Plenum, New York

Amblard, P.O. and Michel, O.J. 2013. The relation between Granger causality and directed information theory: a review. *Entropy* **15**, 113–143.

Arenas, F. and Vaz-Pinto, F. 2014. Marine algae as carbon sinks and allies to combat global warming. In: *Marine Algae*: ( ed. Pereira, L. and Neto, J. M.). pp. 178-194. CRC Press, Boca Raton, Florida.

Åström, K.J. and Murray, R.M. 2008. *Feedback Systems: An Introduction for Scientists and Engineers*. Princeton University Press, Princeton.

Bacastow, R.B. 1976. Modulation of atmospheric carbon dioxide by the southern oscillation, *Nature*, **261**, 116–118.

Ban-Weiss, G.A., Bala, G., Cao, L., Pongratz, J. and Caldeira, K. 2011. Climate forcing and response to idealized changes in surface latent and sensible heat. *Environ. Res. Lett.* **6**, 034032–034038. doi:10.1088/1748-9326/6/3/034032

Bar-On, Y.M., Phillips, R. and Milo, R. 2018. The biomass distribution on Earth. *Proc. Natl. Acad. Sci. U.S.A*. **115**, 6506–6511.

Bishop, O. 2007. *Electronics: Circuits and Systems*. Routledge, London.

BozorgMagham, A.E., Motesharrei, S., Penny, S.G. and Kalnay, E. 2015. Causality analysis: identifying the leading element in a coupled dynamical system. *PLoS One* **10**, e0131226. doi:10.1371/journal.pone.0131226.

Clarke, J. and Mirza, S.A. 2006. Comparison of some common methods of detecting Granger non-causality. *Journal of Statistical Computation and Simulation* **76**, 207-231.

Elliott, G., Rothenberg, T.J. and Stock, J.H.1996. Efficient tests for an autoregressive unit root, *Econometrica*, **64**, 813–836.

Field, C.B., Behrenfeld, M.J., Randerson, J.T. and Falkowski, P.G. 1998. Primary production of the biosphere: integrating terrestrial and oceanic components. *Science* **281**, 237–240.

Franzke, C. 2012 Nonlinear Trends, Long-Range Dependence, and Climate Noise Properties of Surface Temperature *J. Climate* **25**, 4172–4183. https://doi.org/10.1175/JCLI-D-11-00293.1

Freeman, E., Woodruff, S.D., Worley, S.J., Lubker, S.J., Kent, E.C. et al. 2016. ICOADS Release 3.0: a major update to the historical marine climate record. Int. J. Climatol. **37**, 2211–2232. doi:10.1002/joc.4775. Data sourced from Climate Explorer (last accessed, March 15, 2020). https://climexp.knmi.nl/select.cgi?id=658304c13648f3f6b994afdc81cc5573&field=coads_wspd

Geweke J. 1982. Measurement of linear dependence and feedback between multiple time series. *J. Am. Stat. Assoc.* 77, 304-13.

Gnu Regression, Econometrics and Time-series Library (GRETL) (available from: http://gretl.sourceforge.net/, accessed 17 June 2019)

Gordon, H., Sengupta, K., Rap, A., Duplissy, J., Frege, C., Williamson, C. et al. 2016. Reduced anthropogenic aerosol radiative forcing caused by biogenic new particle formation. *Proc. Natl. Acad. Sci. U. S. A*. **113**, 12053-12058.

Granger, C.W.J. 1969. Investigating causal relations by econometric models and cross-spectral methods, *Econometrica* **37**, 424–438.

Greene, W.H. 2012. *Econometric Analysis*, 7th Edn., Prentice Hall, Boston.

Harris, I., Jones, P.D., Osborn, T.J. and Lister, D.H. 2014. Updated high-resolution grids of monthly climatic observations – the CRU TS3.10 Dataset. *Int. J. Climatol*., **34**: 623–642. doi: 10.1002/joc.3711(last accessed, March 15, 2020) href=http://climexp.knmi.nl/select.cgi?field=cru4_cld_25>climexp.knmi.nl/select.cgi?field=cru4_cld_25</a>

Harvey, D.I., Leybourne, S.J. and Taylor, A.M.R. 2008. Testing for unit roots and the impact of quadratic trends with an application to relative primary commodity prices. Discussion Paper No. 08/04, Granger Centre for Time Series Econometrics, University of Nottingham.

Hesse, W., Moller, E., Arnold, M. and Schack, B. 2003. The use of time-variant EEG Granger causality for inspecting directed interdependencies of neural assemblies. *J. Neurosci. Methods* **124**, 27-44.

Hlinka, J., Jajcay, N., Hartman, D. and Paluˇs, M. 2017. Smooth information flow in temperature climate network reflects mass transport. *Chaos: An Interdisciplinary Journal of Nonlinear Science* **27**, 035811.

IHS EViews. 2017. EViews 9.5, IHS Global Inc., Irvine, CA. Online at: http://www.eviews.com/download/download.shtml (last accessed 20 August 2018)

Imbers, J., Lopez, A., Huntingford, C. and Allen, M. R. 2013. Testing the robustness of the anthropogenic climate change detection statements using different empirical models. *J. Geophys. Res.- Atmos.* **118**, 3192–3199.

IPCC. 2007. *Climate Change 2007: Synthesis Report.* Cambridge University Press, Cambridge.

IPCC. 2013. *Climate Change 2013: The Physical Science Basis. Contribution of Working Group I to the Fifth Assessment Report of the Intergovernmental Panel on Climate Change*. Cambridge University Press, Cambridge.

Janjua ,P.Z., Samad, G. and Khan, N. 2014. Climate change and wheat production in Pakistan: an autoregressive distributed lag approach. *NJAS Wagening J Life Sci* **68**, 13–19. https://doi.org/10.1016/j.njas.2013.11.002

Kaufmann, R.K. and Stern, D.I. 1997. Evidence for human influence on climate from hemispheric temperature relations. *Nature*, **38**8,39–44.

Keeling, R.F., Piper, S.C., Bollenbacher, A.F. and Walker, S.J. 2009. Atmospheric CO2 values (ppmv) derived from in situ air samples collected at Mauna Loa, Hawaii, USA Carbon Dioxide Research Group, Scripps Institution of Oceanography (SIO), University of California, La Jolla, California USA 92093-0444, available at: http://cdiac.ornl.gov/ftp/trends/CO2/maunaloa.CO2 (last accessed 15 November 2019).

Lee, H.-T., Gruber, A., Ellingson, R.G. and Laszlo, I. 2007. Development of the HIRS outgoing longwave radiation climate dataset, *J. Atmos. Oceanic Technol.*, **24**, 2029–2047. doi:10.1175/2007JTECHA989.1 OLR data from climexp.knmi.nl/select.cgi?umd_olr   Last accessed 15 March 2020.

Leggett, L.M.W. and Ball, D.A. 2015. Granger causality from changes in level of atmospheric CO2 to global surface temperature and the El Nino–Southern Oscillation, and a candidate mechanism in global photosynthesis. *Atmos. Chem. Phys*. **15**, 11571–11592. doi:10.5194/acp-15-11571-2015

Leggett, L.M.W. and Ball, D.A. 2018. Evidence that global evapotranspiration makes a substantial contribution to the global atmospheric temperature slowdown *Theoretical and Applied Climatology*   https://doi.org/10.1007/s00704-018-2387-7

Leggett, L.M.W. and Ball, D.A. 2020. Observational evidence that a feedback control system with proportional-integral-derivative characteristics is operating on atmospheric surface temperature at global scale, *Tellus A: Dynamic Meteorology and Oceanography*, **72**, 1-14. doi: 10.1080/16000870.2020.1717268

Levitus, S., Antonov, J.I., Boyer, T.P., Baranova, O.K., Garcia, H.E., Locarnini, R.A., Mishonov, A.V., Reagan, J.R., Seidov, D., Yarosh, E.S. and Zweng, M.M. 2012. World ocean heat content and thermosteric sea level change (0–2000 m), 1955–2010. *Geophys Res Lett* **39**:L10603. Ocean heat data used available at https://www.nodc.noaa.gov/OC5/3M%5fHEAT%5fCONTENT/index.html, last accessed 11 December 2019

Lovelock, J.E. and Margulis, L. 1974. Atmospheric homeostasis by and for the biosphere: the Gaia hypothesis. *Tellus Series A* **26**, 2–10.

Morice, C.P., Kennedy, J.J., Rayner, N.A. and Jones, P.D. 2012. Quantifying uncertainties in global and regional temperature change using an ensemble of observational estimates: the HadCRUT4 dataset. *J. Geophys. Res*., **117**, D08101. doi:10.1029/2011JD017187, 2012. HadCRUT4 data used available at: http://www.metoffice.gov.uk/hadobs/hadcrut4/data/current/time_series/HadCRUT.4.4.0.0.monthly_ns_avg.txt, last access: 12 September 2015

Mudelsee, M. 2010. *Climate Time Series Analysis*, Springer, Switzerland.

Pasini, A., Racca, P., Amendola, S., Cartocci, G. and Cassardo, C. 2017. Attribution of recent temperature behaviour reassessed by a neural-network method. *Sci. Rep.* **7**, 17681. https://doi.org/10. 1038/s41598-017-18011-8

Pesaran, M.H., Shin, Y. and Smith R.J. 2001. Bounds testing approaches to the analysis of level relationships. *J. Appl. Econ.* **16**, 289–326. https://doi.org/10.1002/jae.616

Pfaff, B., Zivot, E. and Stigler, M. 2016. Package urca. Unit root and cointegration tests for time series data. https://cran.rproject.org/web/packages/urca/urca.pdf (accessed August 2019)

R Development Core Team. 2009. R: A Language and Environment for Statistical Computing. R Foundation for Statistical Computing, Vienna, Austria. Downloaded from https://mirror.its.sfu.ca/mirror/CRAN/ (accessed September 2019).

Raju, M., Saikia, L.C. and Sinha, N. 2016. Automatic generation control of a multi-area system using ant lion optimizer algorithm based PID plus second order derivative controller. *Int. J. Electr. Power Energy Syst.* **80**, 52e63.

Richards, F. and Arkin, P. 1981. On the relationship between satellite observed cloud cover and precipitation. *Mon. Wea. Rev*., **109,** 1081-1093. doi:10.1175/1520-0493(1981) 109,1081:OTRBSO.2.0.CO;2.

Sahib, M.A. 2015. A novel optimal PID plus second order derivative controller for AVR system. *Eng. Sci. Technol. Int. J*. **18**, 194-206.

Sarmiento, J.L. and Gruber, N. 2006: *Ocean Biogeochemical Dynamics*. Princeton University Press, Princeton, New Jersey.

Sato M., Hansen J.E., McCormick M.P. and Pollack J.B. 1993. Stratospheric aerosol optical depths, 1850–1990. *J. Geophys. Res*. **98**, 22987–22994. Data used available at: http://data.giss.nasa.gov/modelforce/strataer/tau.line_2012.12.txt ( accessed 8 July 2019).

Schneider, S.H., Root, T.L. and Mastrandrea, M.D. 2011. *Encyclopedia of Climate and Weather*. Oxford University Press, Oxford. DOI:10.1093/acref/9780199765324.001.000

Schmidt, P. and Phillips, P.C.B. 1992. LM test for a unit root in the presence of deterministic trends. *Oxford Bulletin of Economics and Statistics* **54**, 257–287. doi:10.1111/j.1468-0084.1992.tb00002.x

Sheil, D.F. 2018. Forests, atmospheric water and an uncertain future: the new biology of the global water cycle. *For. Ecosyst*. **5**, 19. http://dx.doi.org/10.1186/s40663-018-0138-y.

Sims, C.A. 1980. Macroeconomics and reality. *Econometrica*. **48**, 1-48.

Stern, D.I. and Kaufmann R.K. 2014. Anthropogenic and natural causes of climate change. *Clim. Chang*. **122**, 257–269. https://doi.org/10.1007/s10584-013-1007-x

Stock, J.H. and Watson, M.W. 2001. Vector autoregressions. *Journal of Economic Perspectives*, **15**, 101-116.

Sugihara, G., May, R., Ye, H., Hsieh, C-H., Deyle, E., Fogarty, M. and Munch, S. 2012. Detecting causality in complex ecosystems. *Science* **338**, 496–500.

Taylor, K.E., Stouffer, R.J. and Meehl, G.A. 2012. An overview of CMIP5 and the experiment design, *B. Am. Meteorol. Soc*. **93**, 485–498, doi:10.1175/BAMS-D-11-00094.1. CMIP5 data used available at: http://climexp.knmi.nl/data/icmip5_tas_Amon_modmean_rcp45_0-360E_-90-90N_n_+++a.txt, last access: 3 February 2020

Terrell, C. 2019. *Predictions in Time Series using Regression Models.* ED-Tech Press, Waltham Abbey Essex.

Toda, H.Y. and Yamamoto, T. 1995. Statistical inferences in vector autoregressions with possibly integrated processes. *Journal of Econometrics*, **66**, 225-250.

Triacca, U. 2005. Is Granger causality analysis appropriate to investigate the relationship between atmospheric concentration of carbon dioxide and global surface air temperature? *Theor. Appl. Climatol.* **81**, 133–135.

Triacca, U., Attanasio, A. and Pasini, A. 2013. Anthropogenic global warming hypothesis: testing its robustness by Granger causality analysis. *Environmetrics* **24**, 260–268.

Troup, A.J. 1965. The Southern Oscillation, *Q. J. Roy. Meteor. Soc*. **91**, 490–506, 1965. SOI data used available at: https://www.longpaddock.qld.gov.au/seasonalclimateoutlook/southernoscillationindex/soidatafiles/MonthlySOI1887-1989Base.txt, last access: 25 February 2020.

Tucker, C.J., Pinzon, J.E., Brown, M.E., Slayback, D., Pak, E.W., Mahoney, R., Vermote, E. and El Saleous, N. 2005. An extended AVHRR 8-km NDVI data set compatible with MODIS and SPOT vegetation. *Int. J. Remote Sens*. **26**, 4485–5598.

Walcek, C.J. 1994. Cloud cover and its relationship to relative humidity during a springtime mid-latitude cyclone. *Mon. Weather Rev*. **122**, 1021–1035. https://doi.org/10.1175/1520-0493122<1021:CCAIRT>2.0.CO;2

Wilks, D.S. 2011. *Statistical Methods in the Atmospheric Sciences: an Introduction.* Academic Press, London.

Wunsch, C. 2002. What is the thermohaline circulation? *Science* **298**, 1179–1180.

Yan, X.-H., Boyer, T., Trenberth, K., Karl, T.R., Xie, S.-P., Nieves, V., Tung, K.-K. and Roemmich, D. 2016. The global warming hiatus: slowdown or redistribution? *Earth's Future.* **4**, 472–482. https://doi.org/10.1002/2016EF000417

.